\documentclass[sigconf,nonacm]{acmart}

\usepackage{booktabs}
\usepackage{array}
\usepackage{float}
\setcopyright{none}
\hypersetup{pdfauthor={Heng Yao, Hanwei Zhu}}

\begin{document}
\title[Where Beam Search Loses Top-Ranked Items]{Where Beam Search Loses Top-Ranked Items:\\First-Step Coverage in Generative Recommendation}

\author{Heng Yao}
\authornote{Corresponding authors: Heng Yao (\href{mailto:yaoheng.cs@qq.com}{\nolinkurl{yaoheng.cs@qq.com}}) and Hanwei Zhu (\href{mailto:zhuhanweiowen@gmail.com}{\nolinkurl{zhuhanweiowen@gmail.com}}).}
\affiliation{\institution{Ant Group}\country{}}

\author{Hanwei Zhu}
\authornotemark[1]
\affiliation{\institution{Independent Researcher}\country{}}
\renewcommand{\shortauthors}{Yao and Zhu}

\begin{abstract}
Generative recommendation can reproduce a wider beam's ranked output with roughly half the prefix-capacity budget by reallocating retention capacity across semantic-ID positions. We identify a \emph{first-step coverage bottleneck}: across six full-test settings of the TIGER generative recommender spanning three domains and two identifier constructions, every pruned exact top-10 path is first lost at the initial decoding step. An exact-catalog reference isolates search losses and guides a controlled budget intervention. Moving width to the first step raises Search Recall@10 by 0.96--10.90 percentage points at Beam-50's unchanged budget. Across all 77,373 test users with residual-quantized variational autoencoder (RQ-VAE) tokenization, the reallocated schedule reproduces Beam-100's ordered top-10 at nominal budgets of 151 rather than 301, without retraining or changing identifiers or scores. Matched-output timing saves 6.35--6.39\% of decoder latency; profiling identifies the execution costs that limit the translation from reduced search work to runtime savings. Larger-candidate experiments reveal how output size changes retention requirements. These findings make budget placement an actionable serving decision: locate ranking loss, redirect search capacity, and preserve downstream inputs before increasing uniform beam width.
\end{abstract}

\maketitle

\section{Introduction}
A deployed generative recommender must recover its trained ranking within a serving budget. Industrial systems co-design architecture, learning, and beam serving \cite{gr4ad}. Uniform widening spends additional work at every semantic identifier (SID) position, although early steps retain item groups and later steps distinguish items within them. We ask: \emph{can loss localization identify a cheaper allocation that preserves a wider beam's ranked output?}

\begin{figure}[htbp]
\centering
\includegraphics[width=\columnwidth]{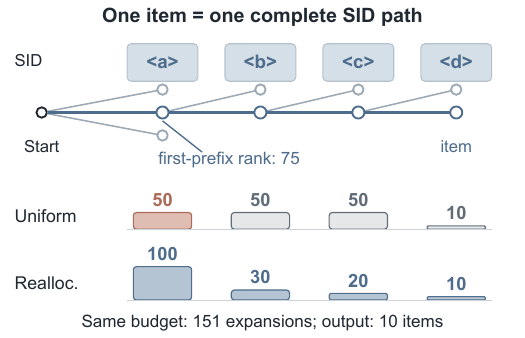}
\caption{SID paths and width reallocation. Tokens represent numeric codes; bar heights share a width scale. Illustrative rank 75 survives width 100; both schedules have budget 151 and return ten items.}
\label{fig:sid-budget}
\end{figure}

Beam search prunes partial identifier paths before complete item scores are available, removing every item beneath a discarded prefix. Figure~\ref{fig:sid-budget} illustrates equal-budget reallocation. We connect \textbf{loss location, recovery through capacity placement, and output-preserving cost reduction}.

\subsection{From Label Loss to Search Loss}
\begin{figure}[htbp]
\centering
\includegraphics[width=0.90\columnwidth]{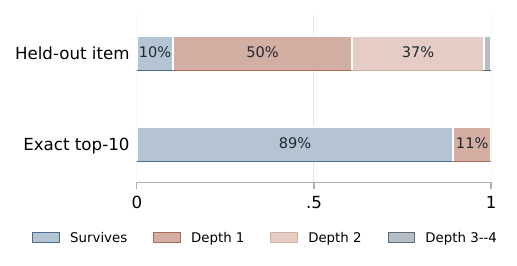}
\caption{Target-path loss spans depths; exact top-10 loss concentrates at step one. Beauty seed 43, legal Beam-50: bars partition held-out targets and exact top-10 paths into survival and first-prune depths.}
\label{fig:pruning}
\end{figure}

Label recall mixes model and search errors. Figure~\ref{fig:pruning} separates them on Beauty seed 43: Beam-50 first removes about 50\% of 22,363 held-out targets at token one and 37\% at token two, while 89.34\% of exact top-10 paths survive and every pruned reference path is first lost at token one. This is a \emph{first-step coverage bottleneck}. Prefix Alignment with Temporal History (PATH) studies target survival \cite{path}; Direct Generative Recommendation (DiGR) develops item ranking \cite{digr}. We test frozen-model search recovery and output-preserving replacement.

Our contributions are:
\begin{itemize}
\item \textbf{An exact-ranked first-step bottleneck.} Every pruned exact top-10 path is first lost at step one across six full-test settings, locating recoverable loss under fixed models.
\item \textbf{A controlled recovery mechanism.} Equal-budget positional controls verify recovery through first-step allocation, raising Search Recall@10 by 0.96--10.90 percentage points.
\item \textbf{Output-preserving serving efficiency.} All 77,373 RQ-VAE test users retain Beam-100's ordered top-10 at roughly half the prefix-capacity budget and 6.35--6.39\% lower measured latency. 
\end{itemize}

\subsection{Relation to Prior Work}
Autoregressive entity retrieval generates names \cite{genre}; a differentiable search index generates document identifiers \cite{dsi}, and a neural corpus indexer adds semantic identifiers and prefix-aware decoding \cite{nci}. The TIGER generative recommender uses semantic identifiers \cite{tiger}. Collaborative-semantic alignment \cite{lcrec}, learnable item tokenization \cite{letter}, and hybrid dense/generative retrieval \cite{liger} change representations or retrieval models; we hold both fixed.

\textbf{PATH: target-path retention.} PATH supervises joint early prefixes and calibrates completed rankings; first-token survival alone need not preserve a target's full path \cite{path}. We locate losses in the frozen model's exact top-10 and test equal-budget positional controls.

\textbf{DiGR: direct item ranking.} DiGR compares beam and exhaustive rankings, then learns item-level scoring \cite{digr}. We keep the model, identifiers, and scores fixed and recover identical ordered outputs with a cheaper decoder.

Adaptive prefix-aware optimization targets vulnerable training prefixes \cite{apao}; item-supported decoding guides prefixes with item rankings \cite{isd}. Advertising systems widen later beams for value \cite{gr4ad}. Adaptive translation beams reduce work \cite{beam}, while exact inference distinguishes model and search errors \cite{searcherrors}. Value-guided Sampling and Tree-structured Advantage Reinforcement (V-STAR) learns value-guided decoding with reinforcement learning \cite{vstar}; we preserve likelihood scores and recover a frozen ranking.

\section{Search Loss and Budget Reallocation}
\subsection{Ranking Reference and Decoding State}
Let $\mathcal U$ be the evaluated user contexts, $\mathcal I$ the legal catalog, and $\theta$ the fixed model parameters. Each item $i$ has a unique length-$L$ SID $z(i)$, including its collision-resolution token. Its complete score is
\begin{equation}\label{eq:score}
s_\theta(i\mid u)=\sum_{t=1}^{L}\log p_\theta(z_t(i)\mid u,z_{<t}(i)).
\end{equation}
Exhaustive scoring gives ordered top-$K$ list $\pi^\star_K(u)$ and its item set $T^K_\theta(u)$. A width schedule $b=(b_1,\ldots,b_L)$ retains at most $b_t$ prefixes at step $t$, with $b_L\ge K$. Let $P_0=\{\emptyset\}$ and $\mathrm{Ext}_{\mathcal I}$ extend retained prefixes by one legal trie token. Beam selection is
\begin{equation}\label{eq:beam}
P_t(u;b)=\operatorname{Top}\!\left(\mathrm{Ext}_{\mathcal I}(P_{t-1}(u;b));b_t\right),
\end{equation}
using Eq.~\ref{eq:score} truncated at $t$. Prefixes can rank differently from their best complete items because the unobserved continuation probabilities differ. The best $K$ completions define ordered output $\pi_K(u;b)$ and item set $B^K(u;b)$. Exact ties use item index and beam ties use SID order; complete-item and prefix-score ordering are checked numerically.

\subsection{Search Fidelity and First-Prune Depth}
For a held-out next item $y_u$, Search Recall (SR) and label Recall are
\begin{equation}\label{eq:metrics}
\begin{aligned}
\mathrm{SR@K}(u;b)&=\frac{|T^K_\theta(u)\cap B^K(u;b)|}{K},\\
\mathrm{Recall@K}(u;b)&=\mathbb{1}\{y_u\in B^K(u;b)\}.
\end{aligned}
\end{equation}
Reported scores average over $\mathcal U$. Search Recall measures fidelity to the frozen ranking; label Recall measures predictive quality. For each reference item, the first-prune depth is
\begin{equation}\label{eq:prune}
d_{b}(u,i)=\min\{t\in[1,L]:z_{\le t}(i)\notin P_t(u;b)\},
\end{equation}
with $d_{b}=0$ when the set is empty (complete survival). Each path contributes once. With identical scores and tie order, a surviving reference item remains top-$K$ among retained candidates. The reference is computed offline.

\subsection{Budget Constraint and Output Contract}
The nominal parent-prefix capacity budget, distinct from elapsed time, is
\begin{equation}\label{eq:budget}
C(b)=1+\sum_{t=1}^{L-1}b_t.
\end{equation}
The initial prefix contributes one; final completions are not expanded. This is a nominal capacity budget; trie exhaustion can lower actual expansions. At fixed model, catalog, and $K$, budget allocation seeks the largest mean Search Recall@K subject to $C(b)\le C_0$ and $b_L\ge K$. First-loss traces guide the schedules tested at budget $C_0$.

For $L=4$, Beam-50/100 use internal width $W=50/100$, giving $C=151/301$. Front-151, $[100,30,20,10]$, moves width to step one at $C_0=151$; middle and late controls use $[30,100,20,10]$ and $[30,20,100,10]$. Front-211 uses $[140,40,30,10]$. Quality runs take the best $K$ from uniform beams; matched timing fixes $b_L=K=10$ for every schedule.

Relative to reference schedule $b^{\rm ref}$, output preservation requires
\begin{equation}\label{eq:contract}
\pi_K(u;b)=\pi_K(u;b^{\rm ref})\quad\forall u\in\mathcal U.
\end{equation}
Equality includes order. We measure synchronized decoder time $\tau(u;b)$ separately from $C$; paired savings are the median of $100[1-\tau(u;b)/\tau(u;b^{\rm ref})]$. For larger-output experiments, $R=|\{z_1(i):i\in\mathcal I\}|$ denotes the catalog root count and $W$ the uniform internal width.

\section{Experiments and Analysis}
We test first-step loss localization, equal-budget recovery, Top-5 recovery, capacity efficiency, and output-preserving latency savings in sequence, then examine output-size sensitivity.

\subsection{Experimental Setup}
We use the GRID toolkit's TIGER implementation with a Text-to-Text Transfer Transformer \cite{grid,tiger,t5} and public Amazon preprocessing \cite{amazon}. Beauty, Sports, and Toys contain 22,363, 35,598, and 19,412 test users and 12,101, 18,357, and 11,924 items. Each domain trains TIGER with residual K-means (RQ-KM) or RQ-VAE \cite{rqvae} tokenization from the same 768-dimensional item vectors; all full identifiers are unique. Sports RQ-VAE trains for 8,000 steps and validation Recall@5 selects step 7,400. Seen items are retained to match GRID.

Main schedules arose from Beauty test exploration and were fixed for Sports/Toys; Beauty is the development domain. Exact scores (Eq.~\ref{eq:score}), legal beams, and schedules are paired per user, with cached scoring checked against uncached scoring. Error bars show 95\% paired user-bootstrap intervals for fixed schedules. Timing uses 512 users per model, excludes eight warm-ups, and retains ten completions for every schedule. Measurements use PPU-ZW810E, Torch 2.6's CUDA-compatible backend, 32-bit floating point with TensorFloat-32 disabled, and one context per call. Decoder order is randomized and work synchronized; encoding and initial scorer preparation are outside the timed call.

\subsection{First-Step Allocation Recovers Model Rankings}
\textbf{First-step coverage is the shared search bottleneck.} The motivating trace in Figure~\ref{fig:pruning} extends to all six settings: every exact top-10 path pruned by either legal beam has $d_{b}(u,i)=1$ (Eq.~\ref{eq:prune}). Early prefixes group multiple items, so first-step capacity preserves branches whose strong continuations are not yet visible.\footnote{A boundary check reproduces 18 near-tied score-order events in 11 users at $K=5/10$, with gaps below 0.00002; reference-score reranking resolves all.}

Table~\ref{tab:main} tests this recovery at $C_0=151$ (Eq.~\ref{eq:budget}). Front-151 improves Search Recall@10 in every setting at Beam-50's capacity budget. Sports RQ-VAE gains 10.90 percentage points. Toys RQ-VAE gains less because Beam-50 already recovers 99.03\% of the reference.

\begin{table}[htbp]
\caption{First-step reallocation improves full-test Search Recall@10 at fixed budget. Beam-50/front-151 have budget 151; Beam-100 has 301. Gain is front-151 minus Beam-50.}
\label{tab:main}
\centering\footnotesize
\setlength{\tabcolsep}{2.4pt}
\begin{tabular*}{\columnwidth}{@{\extracolsep{\fill}}lrrrrr@{}}
\toprule
Domain / Tokenizer & Users & Beam-50 & Front-151 & Gain & Beam-100 \\
\midrule
Beauty / RQ-KM & 22,363 & 0.8934 & \textbf{0.9813} & \textbf{+0.0879} & 0.9814 \\
Sports / RQ-KM & 35,598 & 0.8947 & \textbf{0.9882} & \textbf{+0.0935} & 0.9898 \\
Toys / RQ-KM & 19,412 & 0.9856 & \textbf{0.9999} & \textbf{+0.0143} & 0.9999 \\
Beauty / RQ-VAE & 22,363 & 0.9466 & \textbf{0.9950} & \textbf{+0.0483} & 0.9950 \\
Sports / RQ-VAE & 35,598 & 0.8839 & \textbf{0.9929} & \textbf{+0.1090} & 0.9929 \\
Toys / RQ-VAE & 19,412 & 0.9903 & \textbf{0.9999} & \textbf{+0.0096} & 0.9999 \\
\bottomrule
\end{tabular*}
\end{table}

A positional control holds the Sports RQ-VAE checkpoint, all 35,598 test users, and budget $C_0=151$ fixed. Front-151 recovers 99.29\% of the model's exact top-10 items, versus 74.92\% under either middle or late allocation: a gap of 24.37 percentage points from changing width positions alone (Section~2.3). \textbf{Budget placement improves recovery without increasing the prefix-capacity budget.} Later selection refines retained groups but cannot recover discarded roots.

\begin{figure}[htbp]
\centering
\includegraphics[width=0.90\columnwidth]{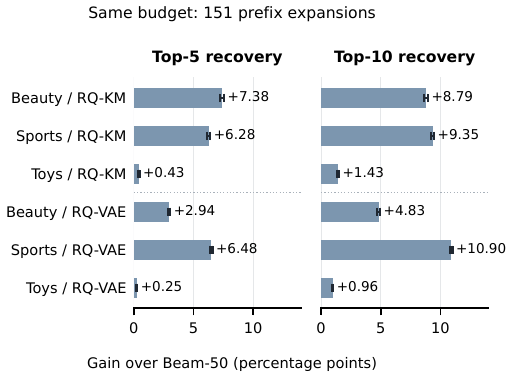}
\caption{Fixed-budget recovery persists at Top-5 and Top-10. Front-151 gains over Beam-50 (percentage points; budget 151 each).}
\label{fig:krobust}
\end{figure}

\textbf{Recovery includes the highest-ranked outputs.} Figure~\ref{fig:krobust} extends the fixed-budget comparison from Top-10 to Top-5, with positive gains in all six settings. Sports RQ-VAE gains 6.48 percentage points at Top-5 and 10.90 at Top-10. The benefit reaches the first five recommendations, not just ranks six to ten.

\begin{figure}[htbp]
\centering
\includegraphics[width=0.94\columnwidth]{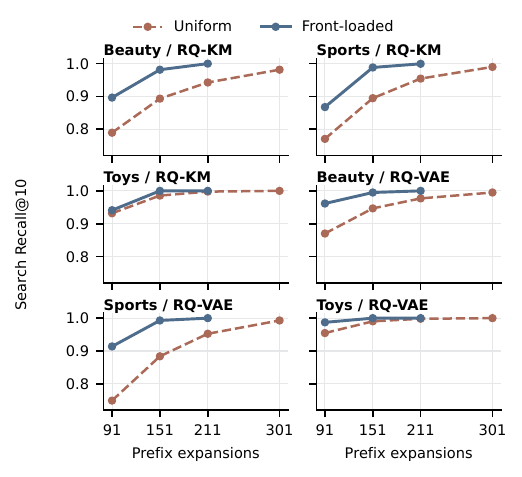}
\caption{Front-loading improves the fidelity--capacity tradeoff. Full-test Search Recall versus nominal prefix capacity $C$.}
\label{fig:pareto}
\end{figure}

\textbf{Better placement improves the fidelity--capacity tradeoff.} Figure~\ref{fig:pareto} then varies the budget: front-loaded schedules reach high fidelity with less capacity than uniform widening. Front-211 exceeds Beam-100's Search Recall@10 in all six settings using $211$ rather than $301$ units of capacity, approximately 30\% less. We next test whether this capacity advantage yields lower latency.

\subsection{Identical Outputs with Lower Decoder Cost}
Full-test ordered-list comparison verifies that front-151 and Beam-100 return identical top-10 lists for all 77,373 RQ-VAE test users (Eq.~\ref{eq:contract}), at $C=151$ versus $301$. A wider beam's output needs roughly half the prefix-capacity budget. 

Table~\ref{tab:profile} compares front-151 with $[100,100,100,10]$, fixing final capacity at ten. \textbf{Output-preserving decoder savings are 6.35--6.39\%.} Thus, the same ranked output takes about 6\% less total decoder time on top of existing GPU-compatible parallel execution, without retraining or changing scores.

Forward/cache management occupies 76.78--79.63\% of Beam-100 wall time (Table~\ref{tab:profile}). Profiling uses 64 users per setting, three repeats, and randomized instrumented/control order; outputs agree and scores match within $10^{-6}$.

\begin{table}[htbp]
\caption{Reallocation reduces measured latency despite dominant forward/cache cost. Median time (ms), paired saving (\%), and baseline forward/cache share (\%) use separate timing/profiling runs. RQ-VAE full-test ordered outputs match Beam-100.}
\label{tab:profile}
\centering\footnotesize
\setlength{\tabcolsep}{2.4pt}
\begin{tabular*}{\columnwidth}{@{\extracolsep{\fill}}lrrrr@{}}
\toprule
Domain / Tokenizer & Beam-100 & Front-151 & Saved & Forward \\
\midrule
Beauty / RQ-KM & 16.63 & \textbf{15.42} & \textbf{6.61} & \textbf{79.15} \\
Sports / RQ-KM & 17.68 & \textbf{16.21} & \textbf{7.00} & \textbf{77.83} \\
Toys / RQ-KM & 17.40 & \textbf{16.26} & \textbf{6.01} & \textbf{79.63} \\
Beauty / RQ-VAE & 17.45 & \textbf{16.33} & \textbf{6.35} & \textbf{77.32} \\
Sports / RQ-VAE & 17.41 & \textbf{16.27} & \textbf{6.39} & \textbf{76.78} \\
Toys / RQ-VAE & 17.05 & \textbf{15.90} & \textbf{6.35} & \textbf{78.33} \\
\bottomrule
\end{tabular*}
\end{table}

Forward/cache execution dominates baseline decoder time. Narrower later beams reduce candidate and selection work while retaining four forward rounds and their call/cache path. These retained costs limit the whole-decoder speedup. 

Predictive quality changes less. On the same 35,598 Sports RQ-VAE users, target hits rise from 659 under Beam-50 to 677 under front-151: 18 additional hits, with Recall@10 increasing from 0.0185 to 0.0190 (Eq.~\ref{eq:metrics}). Large ranking recovery thus yields a modest target-hit gain: search fidelity and target prediction are distinct objectives. Front-151 improves fidelity over Beam-50 at equal capacity and preserves Beam-100's outputs with lower latency.

\subsection{Candidate-Pool Scaling and Robustness}
\begin{figure}[htbp]
\centering
\includegraphics[width=0.94\columnwidth]{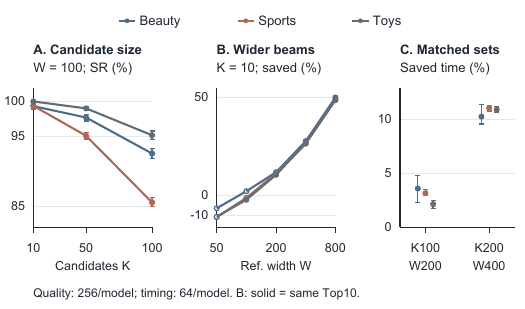}
\caption{Output size changes allocation demands. Exploratory RQ-VAE scaling: 256 quality/64 timing users per model. (A) Candidate sensitivity. (B) Fixed Top10; solid points mean identical ordered outputs. (C) Matched sets at $K/W=100/200$ and $200/400$.}
\label{fig:candidate-scale}
\end{figure}

Figure~\ref{fig:candidate-scale} explores scaling beyond Top-10: schedules were selected after the Sports pilot, with 256 quality and 64 timing users per RQ-VAE model.

Panel A holds internal width at 100: Sports Search Recall falls from about 99\% at $K=10$ to 86\% at $K=100$. Larger requested pools require re-evaluating the schedule.

Panel B holds the output at ten and widens the reference beam. The compact schedule $[R,20,20,10]$ preserves widths 200--800's ordered outputs on all measured users. It saves 10.59--11.89\% versus width 200 and 48.64--49.91\% versus width 800. Additional width becomes removable work when output is unchanged.

For larger outputs, the schedule $[R,\lceil1.5K\rceil,\lfloor1.5K\rfloor,K]$ preserves reference candidate sets on all 256 users per model at $K=100/200$, saving 2.13--3.60\%/10.29--11.09\% against widths 200/400. Candidate size and reference width change together. \textbf{Beam width and requested output size create different allocation demands.} Panel B exposes redundant work at fixed Top10; Panels A/C show larger outputs need more retention. Multi-depth pruning requires width at affected steps; serving objectives also shape advertising schedules \cite{gr4ad}.

Results cover three domains and two SID constructions; offline reference scoring is separate from recurring decoder savings.

\balance
\paragraph{Research artifacts.}
A large language model coding assistant supported scripts, figures, and log inspection. GRID/TIGER and deterministic Python produced the numerical results; retained weight hashes, identifiers, scorer checks, ranked lists, traces, and timing records link each comparison to its evidence.

\section{Conclusion}
Across six TIGER settings, first-loss attribution locates a first-step bottleneck; equal-budget controls and budget sweeps verify recovery through width reallocation. Across all three RQ-VAE models, front-151 reproduces Beam-100's ordered output at roughly half the nominal prefix-capacity budget and 6.35--6.39\% lower measured latency. Profiling and candidate-pool tests explain runtime and capacity demands. \textbf{Width placement improves serving efficiency.}  Locate search loss, redistribute work, and verify output and cost before expanding capacity.
\clearpage
\nobalance
\section*{Ethical Considerations}
The study uses public, preprocessed product-interaction benchmarks and reports aggregate model outputs. It does not infer sensitive attributes or evaluate individual users. Better fidelity to a model's ranking can also faithfully reproduce that model's popularity or representation biases; deployment would require separate quality, fairness, and privacy evaluation. No user-facing intervention or human-subject experiment was conducted.

\bibliographystyle{ACM-Reference-Format}
\bibliography{references}

\end{document}